\documentstyle[masaps,graphicx,html,myprabib,twocolumn]{revtex}

\newcommand{\wavefn}{wavefunction}

\newcommand{\crtot}{Cr$_2$O$_3$}
\newcommand{\fetot}{Fe$_2$O$_3$}
\newcommand{\altot}{Al$_2$O$_3$}
\newcommand{\vtot}{V$_2$O$_3$}

\newcommand{\ai}{\emph{ab initio}}
\newcommand{\df}{density functional}

\newcommand{\lda}{local density approximation}
\newcommand{\lsda}{local spin-density approximation}
\newcommand{\dft}{\df{} theory}

\newcommand{\neel}{N\'eel}

\newcommand{\vps}{pseudopotential}

\newcommand{\DEG}{$^\circ$}

\newcommand{\dftpp}{DFT\raisebox{0.3ex}{\scriptsize ++}}
\newcommand{\bulkpic}[2]{%
\begin{minipage}{1.6in}\begin{center}\setlength{\fboxsep}{1pt}%
\fbox{%
\includegraphics[scale=0.4]{#2}%
}\\%
({#1})
\end{center}\end{minipage}%
}
\newcommand{\slabpic}[2]{%
\begin{minipage}[b]{1.6in}\begin{center}\setlength{\fboxsep}{1pt}%
\fbox{%
\includegraphics[scale=0.40]{#2}\rule[-0.4in]{0em}{2.8in}}\\%
({#1})%
\end{center}\end{minipage}%
}

\title{
\emph{\textbf{Ab initio}} study of magnetic structure and chemical reactivity of
\crtot{}\\and its (0001) surface}

\author{Jason A. Cline$^\dag$ \and Angeliki A. Rigos$^\ddag$ \and Tom\'as A. Arias$^*$}

\address{$^\dag$Energy Laboratory and Department of Chemical Engineering,\\
Massachusetts Institute of Technology, Cambridge, MA 02139.\\
$^\ddag$Department of Chemistry,\\ Merrimack College, North Andover, MA
01845.\\
$^*$Laboratory for Atomic and Solid State Physics,\\
Cornell University, Ithaca, NY 14853.}

\date{Submitted to the Journal of Physical Chemistry B on 6 December, 1999}

\abstract{We present the first \ai{} \dft{} study of the
oxygen-terminated \crtot{} (0001) surface within the local
spin-density approximation (LSDA).  We find that spin plays a critical
role for even the most basic properties of \crtot{} such as the
structure and mechanical response of the bulk material.  The surface
exhibits strong relaxations and changes in electronic and magnetic
structure with important implications for the chemical reactivity and
unusual spin-dependent catalytic activity of the surface.  Unlike the
bulk, the outermost chromium bilayer is ferromagnetically ordered, and
the surface oxygen layer exhibits appreciable net spin polarization in
the opposite sense.  Surprisingly, despite this ferrimagnetic order,
the chemically important states near the Fermi level exhibit
ferromagnetic order and thus favor electronic spin alignment of
species interacting with the surface.  Finally, we also find a high
density of unoccupied electronic surface states available to
participate in the chemical reactivity of the surface.  }

\begin{document}
%------------------------------------------------------------------------------

\maketitle

%------------------------------------------------------------------------------
\section{Introduction}
%------------------------------------------------------------------------------

Chromium oxide is useful as a catalyst in many applications including
internal combustion engine emission technology \cite{Harrison} and the
dehydrogenation of alkanes \cite{Bond,mentasty}.  It is also the
primary constituent of passive films protecting stainless steels and
other high performance industrial alloys
\cite{alstrup,Figueiredo,Ryan}.  Moreover, the surface is used in
industry to catalyze the conversion of para- to
ortho-hydrogen\cite{Ilisca}.  Intriguingly, this conversion has been
observed to decrease by nearly an order of magnitude as the
temperature increases through the \neel{} magnetic ordering
temperature\cite{AriasSelwood}.  That this conversion is so efficient
for a non-magnetic material such as \crtot{} has remained a subject of
discussion for several decades\cite{Ilisca}.  Despite these intriguing
and important phenomena, the surface electronic structure of \crtot{}
is the least studied among the transition metal oxides\cite{henrich}.

Studies of corundum-structure metal oxides to date have neglected the
electronic and magnetic structure of the surfaces of \crtot{}.
Manassidis {\em et al.}\ \cite{Manassidis} were one of the first to
use \ai{} \dft{} (DFT) within the \lda{} (LDA) to study the structure
and energetics of basal plane surfaces of a corundum crystal (\altot).
They found very large surface relaxations which resulted in a
reduction of the surface energy by a factor of two, but, as \altot{}
is non-magnetic, whether transition metal systems exhibit this
behavior remained unexplored.  Veliah {\em et al.}\ \cite{cluster}
employed DFT with both the \lsda{} (LSDA) and non-local LSDA (NLSDA)
to study Cr$_m$O$_n$ clusters, but only up to $m=2, n=3$, too small to
provide insight into the behavior of surfaces.  Classical molecular
dynamics calculations\cite{ssp:rohr_cr2o3_surf} as well as \ai{}
periodic unrestricted Hartree-Fock (UHF) calculations of bulk
\crtot{}\cite{catti} and its surface \cite{ai:rehbein_srl} have been
carried out, but these studies have neglected the oxygen-terminated
surface, which is important in oxygen-rich environments.  Such
surfaces have been studied, but in the related transition-metal
material $\alpha$-\fetot{}.  Using spin-DFT within the generalized
gradient approximation and a full potential linearized augmented plane
wave (FP-LAPW) basis set, Wang et al.\ \cite{dft:fe2o3_scheffler}
studied the $\alpha$-\fetot{} (0001) surface as a function of
increasing oxygen pressure and demonstrated that, indeed, the
oxygen-terminated surface eventually becomes the most stable.

To better understand the unique properties of \crtot{}, we explore the
oxygen-terminated (0001) surface of \crtot{} using the {\em ab initio}
plane-wave pseudopotential approach within LSDA\@.
Section~\ref{sec:comp} briefly reviews our methodology, and
Section~\ref{sec:bulk} gauges the reliability of this approach through
detailed comparisons with experimental information on the lattice
constant, bulk modulus, and atomic arrangement and magnetic structure
of the bulk material.  Finally, in Section~\ref{sec:surf}, we contrast
the atomic, electronic, and magnetic structures of the
surface with those of the bulk.

%------------------------------------------------------------------------------
\section{Methodology} \label{sec:comp}
%------------------------------------------------------------------------------

Our calculations are carried out within the \ai{} \vps{}
density-functional formalism which has been applied successfully in
the past to a wide variety of solid state and surface systems.  (See
\cite{RMP} for a review.)  The present calculations involve three
primary approximations: the local density approximation to density
functional theory, the \vps{} (``frozen-core'') approximation and the
supercell approximation.

To evaluate the efficacy of local density functionals in treating the
present material, we have calculated the properties of bulk \crtot{}
within both LDA and LSDA as parameterized in \cite{PerdewZunger2} and
\cite{PerdewWang}, respectively.  These results
(Section~\ref{sec:bulk}) show that LSDA gives a description far
superior to the LDA and also gives results better than those reported
in the literature for unrestricted Hartree-Fock (UHF)
calculations\cite{catti}.  Accordingly, we employ LSDA exclusively for
the surface calculations below.

To represent the chromium and oxygen ionic cores in this study, we
employ \vps{}s of the non-local Kleinmann-Bylander separable
form~\cite{KleinmanBylander} as generated using the optimized \vps{}
procedure of Rappe et al.~\cite{Rappepots}.  Preliminary calculations
carried out on small atomic clusters indicate that the states of the
argon shell are relatively close in energy to the $2s$ states of
oxygen.  Out of concern for the possible mixing of the argon-shell
electrons of chromium with the oxygen valence shell, we include the
argon-shell electrons explicitly in our calculation as valence
electrons and pseudize only the neon core of chromium in our
pseudopotential.  Convergence tests show that both this \vps{} and
that for oxygen are well-converged at a plane wave cutoff of 80
Rydbergs, which we use for all results reported below.  We find that,
relative to the argon-core \vps{}, the neon-core potential reduces
errors in the band structure relative to the results of LAPW
calculations\cite{ssp:papa} by a factor of two.  

For the bulk calculations described below, we employ the ten-atom
corundum primitive cell of \crtot{}.  For this cell, we sample the
Brillouin zone with an eight k-point
Monkhorst-Pack~\cite{MonkhorstPack} mesh, which folds to four points
under time-reversal symmetry.  This corresponds to a reciprocal-space
sampling density of 0.37~bohr$^{-1}$, which is commonly used to treat
silicon and which we expect to be more than sufficient for this even
more ionically bonded system.

For the surface calculations, we employ the supercell approach with
periodic boundary conditions in all three dimensions, thereby allowing
us to study isolated slabs of Cr$_2$O$_3$ material with two (0001)
surfaces separated by 7~\AA\   of vacuum.  This supercell contains five
bilayers of chromium and six layers of oxygen.  We arrange these to
form a slab with two identical oxygen-terminated surfaces in a unit
cell with the in-plane lattice constant held at the {\em ab initio}
value.  Such slabs were found sufficient for studies of the
corresponding surface of \fetot{}\cite{dft:fe2o3_scheffler} and avoid
the formation of a net electric dipole in the supercell.  For the
surface cells, we use four k-points (which fold to two) in the plane
perpendicular to the $c$-axis of the cell.  This yields a reciprocal
space sampling comparable to that of our bulk calculations.

Within the above approximations, we determine the quantum state of the
system by minimizing the total energy over all possible sets of
orthonormal electronic \wavefn{}s using the analytically-continued
functional approach \cite{RMP,dynamprl}, which has been recently shown
to be nearly optimal in a rigorous mathematical sense\cite{gccg}.  The
calculations are performed within the \dftpp{}
formalism\cite{dirac,RMPwavelet} which makes different physical
descriptions (LDA, LSDA, SIC, Hartree-Fock, etc.) relatively simple to
explore and provides both high portability and good computational
performance.  Finally, when determining structures, we relax the
atomic coordinates until the Hellmann-Feynman forces on the atomic
cores are less than 0.03~eV/\AA\@ in each coordinate direction.

%------------------------------------------------------------------------------
\section{Results and Discussion} \label{sec:prere}
%------------------------------------------------------------------------------

\subsection{Bulk \crtot{}} \label{sec:bulk}

\begin{figure}
\begin{center}
\includegraphics[scale=0.6]{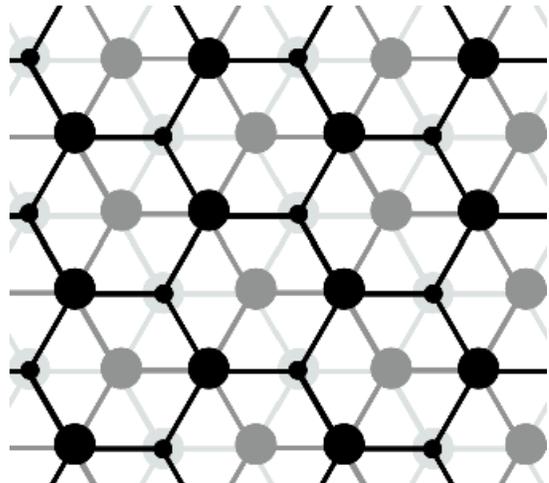}
\caption{Schematic projection of the metallic atoms in the corundum
structure, viewed atop the (0001) surface, showing chromium atoms
only.  The chromium atoms in the structure are colored according depth
and are arranged in a sequence of bilayers with honeycomb structure
(three are shown).  The upper and lower chromium atoms of each bilayer
appear as large and small circles, respectively.}\label{fig:honeycomb}
\end{center}
\end{figure}

The crystal structure of \crtot{} is important for understanding the
relaxation of the internal coordinates of the unit cell, the magnetic
order of the bulk, and the structure of the surface.  \crtot{} assumes
a corundum-type structure (space group $R\bar{3}$~c), which has a
ten-atom primitive cell that is equivalent to a thirty-atom cell on a
hexagonal lattice.  The structure consists of alternating oxygen
layers and chromium bilayers stacked along the $c$-axis of the
hexagonal lattice.  Figure~\ref{fig:honeycomb} shows the arrangement
of the chromium bilayers.  Each bilayer consists of two perfectly
planar triangular lattices arranged so that their combined projection
forms an ideal honeycomb structure whose two-element basis consists of
one atom from each sublayer.  The difference between the
crystallographic parameter ``$z$(Cr)''\cite{Hahn} and one-third
measures the separation of the planes which make up the bilayer.  In
the structure, the chromium bilayers align in an ABC stacking sequence
along the $c$-axis such that atoms from the furthest sublayers of each
bilayer pair align directly along the $c$-axis.  All oxygen atoms in a
layer lie in exactly the same plane and form an approximate triangular
lattice.  The difference between the crystallographic parameter
``$x$(O)''\cite{Hahn} and one-third measures a slight contraction of
those triangles in this lattice which are centered on $c$-axis aligned
pairs of chromium atoms.

It is known experimentally that, unlike \fetot{} and \vtot, both
\altot{} and \crtot{} contract nearly isotropically (within tenths of
a percent) under hydrostatic pressure\cite{bulk2}.  Accordingly, for
our LDA and LSDA calculations, we determined bulk moduli and unit cell
volume by holding the ratio $c/a$ between the hexagonal crystal axes
constrained to the experimental value of
2.7407\cite[p. 469]{LandoltBornsteinvol7}.  At each volume, we
optimized the internal coordinates of the unit cell, and applied a
standard Pulay stress correction\cite{FroyenCohen,GomezDacostaEtAl}
for changes in finite-basis size to the resulting total energy.  The
minimum energy, location of the minimum and the second derivative of
the energy as a function of lattice constant give the cohesive energy,
unit-cell volume and bulk modulus, respectively.

\begin{figure}
\begin{center}
\bulkpic{a}{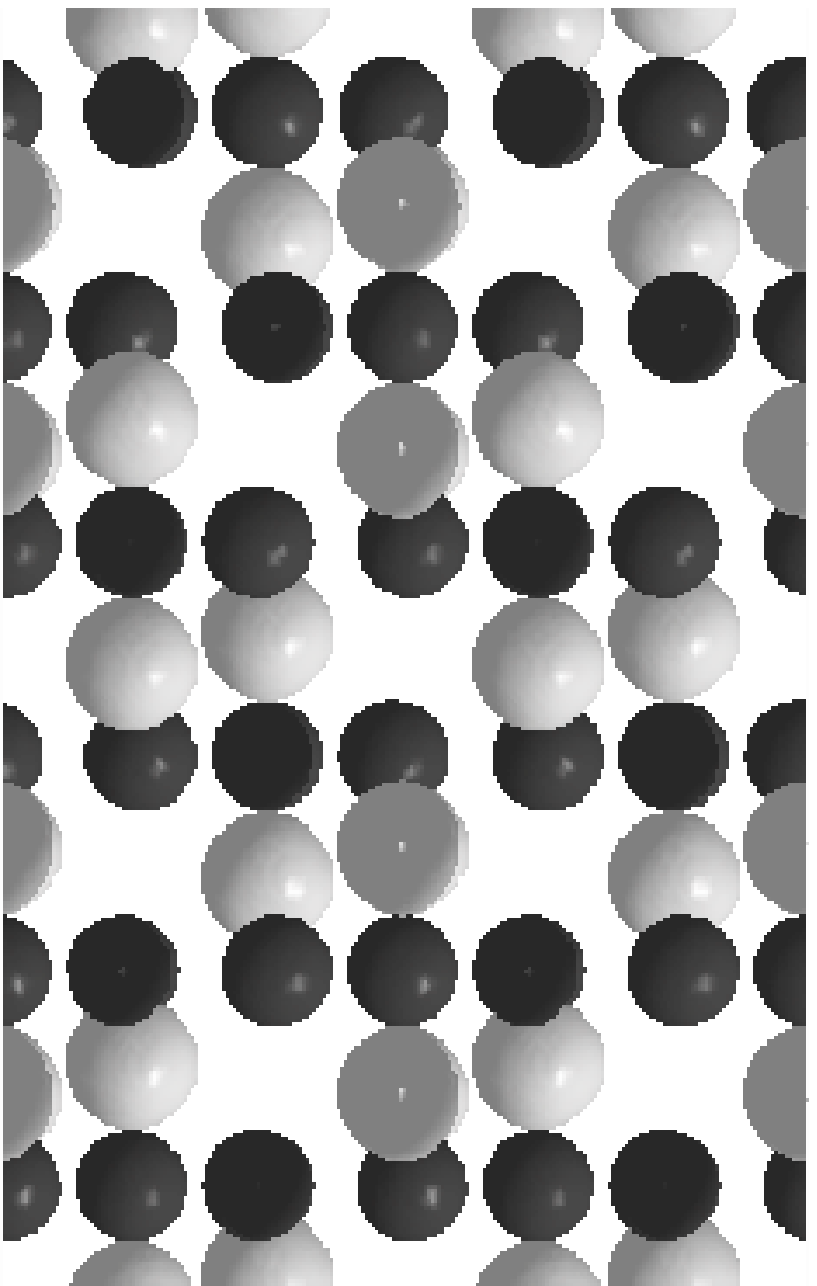}
\bulkpic{b}{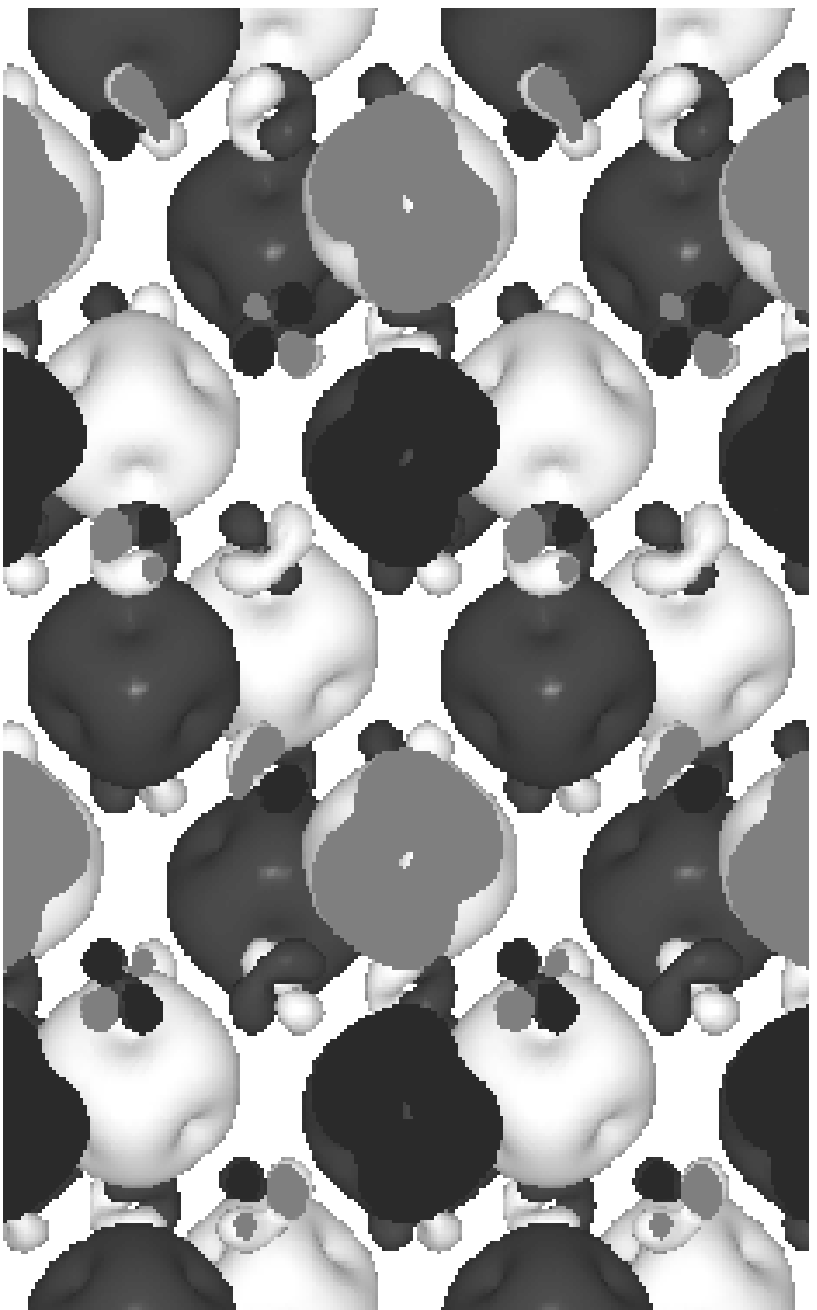}
\end{center}
\caption{LSDA predictions for a thirty atom cell of bulk \crtot{}.
Left panel (a) shows a single isosurface of the predicted total
electron density ($n_\uparrow+n_\downarrow$), with the surface colored
according to proximity to chromium atoms (white) and oxygen atoms
(charcoal).  Right panel (b) shows the spin polarization,
$n_\uparrow-n_\downarrow$, for bulk \crtot{} within LSDA\@ (thirty
atom calculation): contour of net up-spin (white), contour of net
down-spin (charcoal).  For both panels some atoms are cut-off at the
cell boundary, revealing darker shaded interiors.  }
\label{fig:spin_bulk_30}
\end{figure}

Table~\ref{table:bulk_mod} compares our results for bulk modulus, unit
cell volume, cohesive energy (atomization energy to neutral atoms),
and internal cell coordinates with experiment and unrestricted
Hartree-Fock results\cite{catti}.  Whereas the LDA calculations are in
error by about \mbox{-10\%} and 44\% for unit cell volume and bulk
modulus, respectively, the corresponding errors for UHF are +5\% and
+13\% and for LSDA are -1.7\% and within experimental error (5\%),
respectively.  Moreover, in going from LDA to LSDA, the error of the
cohesive energy, determined from Born-Haber cycle data for \crtot{}
\cite{sherman} and the bond strength of molecular oxygen \cite{crc},
reduces from -12\% to -6\%.  (Ref.~\cite{catti} does not provide a
value of the cohesive energy.)

We find that the LDA prediction of the internal coordinates of the
unit cell is in significant error.  We note that $x$(O) for the LDA
structure is nearly one-third, indicating a tendency toward a perfect
triangular lattice in each O layer.  Our LSDA results, on the other
hand, are in much better accord with experiment, showing even somewhat
better agreement than reported for UHF \cite{catti}.  The
uncertainties quoted for $x$(O) and $z$(Cr) in the table represent
small differences in position of equivalent atoms due to numerically
imperfect relaxation.

Given the success of the LSDA description and its clear superiority
over the LDA (and UHF), all results reported below are computed within
LSDA\@.

Next, we turn our attention to the magnetic structure of bulk \crtot.
Figure~\ref{fig:spin_bulk_30}b presents our prediction for the
ground-state spin density.  Consistent with expectations, we find that
the spin polarization in the bulk material concentrates almost
entirely on the chromium atoms (large dimpled contours in the figure).
The spin order is antiferromagnetic among nearest neighbors in each
chromium bilayer, leading to an intra-layer order in which the spin
direction is constant in each sublayer.  Moreover, we find an
inter-layer order such that the spin-up/spin-down sequence along
the $c$-axis is identical for each bilayer.  These results are in
complete accord with the magnetic order inferred from neutron
diffraction experiments\cite{Brockhouse}.  Finally, we find that the
oxygen atoms have lobes of opposite spin-polarization with negligible
net spin.

\begin{figure}
\begin{center}
\includegraphics[scale=0.6]{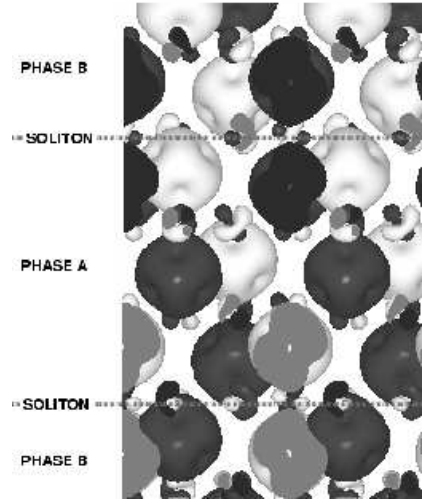}
\end{center}
\caption{Antiferromagnetic soliton in \crtot{} within LSDA\@ (same
conventions as in Figure~\ref{fig:spin_bulk_30}b).  }
\label{fig:spin_bulk_30_sol}
\end{figure}

To confirm that the state shown in Figure~\ref{fig:spin_bulk_30}b as
the ground state within LSDA, we relaxed our initial electronic state
from completely random wavefunctions.  This resulted in a spin state
in which the magnetic order within each bilayer was the same as
described above, suggesting that the antiferromagnetic intra-layer
order is quite stable.  However, in this calculation, the
spin-up/spin-down sequence of the bilayers reversed from one ground
state to the spin-flip related phase.  (See
Figure~\ref{fig:spin_bulk_30_sol}.)  This corresponds to the presence
of two antiferromagnetic solitons in the supercell.  LSDA ascribes a
positive energy to this excitation, indicating that the spin ordering
shown in Figure~\ref{fig:spin_bulk_30}b represents the ground state.
From our results, we extract the first \ai{} estimate for the (0001)
$c$-axis antiferromagnetic coupling constant, $J_{AB}\,\approx
150\,\mathrm{cm^{-1}}$.  Experiments provide highly variable estimates
for this coupling constant from about $J_{AB} \approx
250\,\mathrm{cm^{-1}}$ to $\approx 350\,\mathrm{cm^{-1}}$\cite{poole}.

\begin{figure}
\begin{center}
\bulkpic{a}{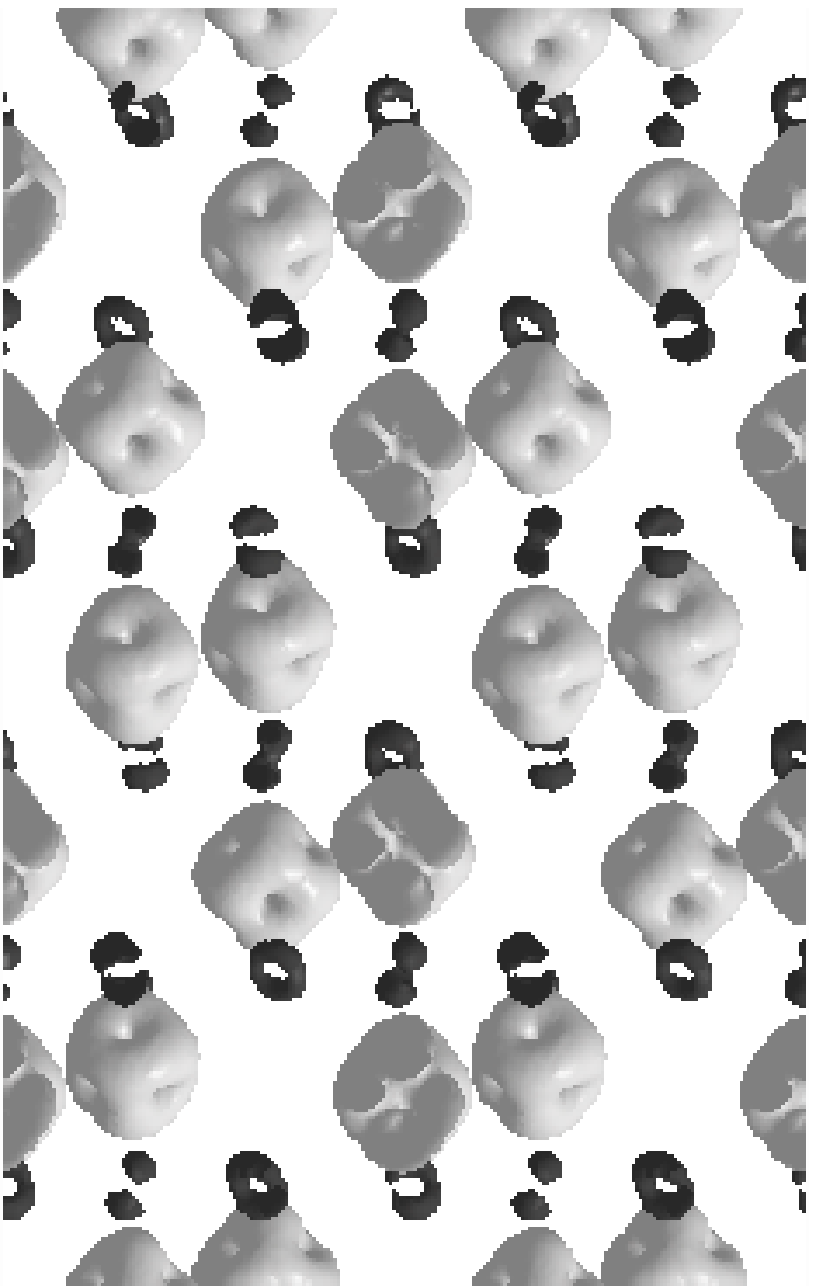}
\bulkpic{b}{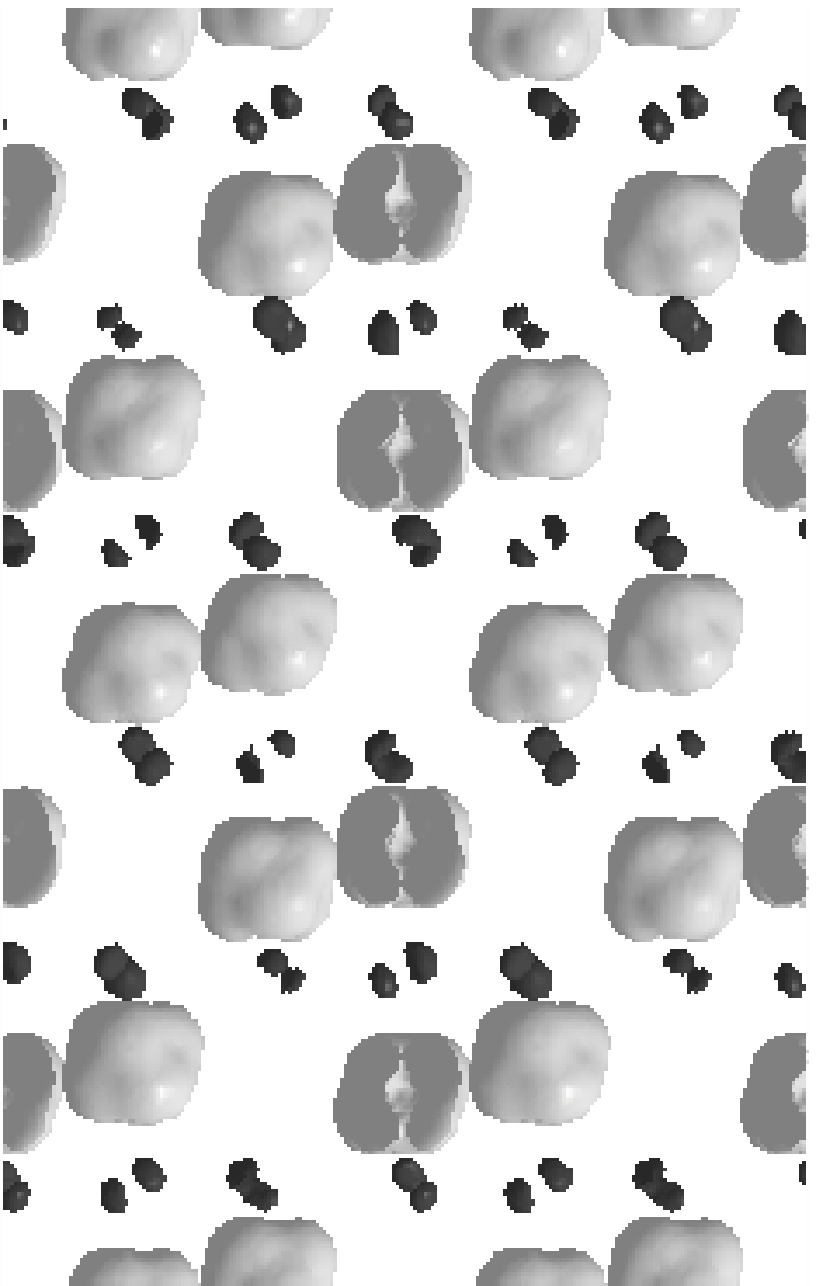}
\ \\
\caption{Local chemical softness map of bulk \crtot{}.  Portrayed are
the HOMOs (a) and LUMOs (b) as calculated within LSDA\@.  The figure
shows a single isosurface of local softness, which is colored
according to proximity to either chromium (white) or oxygen
(charcoal) atoms.
\label{fig:lsda_homolumo_30}
}
\end{center}
\end{figure}

Finally, Figure~\ref{fig:lsda_homolumo_30} shows contour surfaces of
the electron density of the highest occupied molecular orbitals
(HOMOs) and the lowest unoccupied molecular orbitals (LUMOs),
representing the local chemical softness\cite{YangParr}.  Species
which tend to accept or donate electrons tend toward regions with
strong HOMO or LUMO concentration, respectively.  The figure shows
that, in bulk, the chromium atoms interact with either type of
species, whereas the oxygen atoms are relatively inert.

\subsection{Oxygen-terminated (0001) surface} \label{sec:surf}

\begin{figure}
\begin{center}
\slabpic{a}{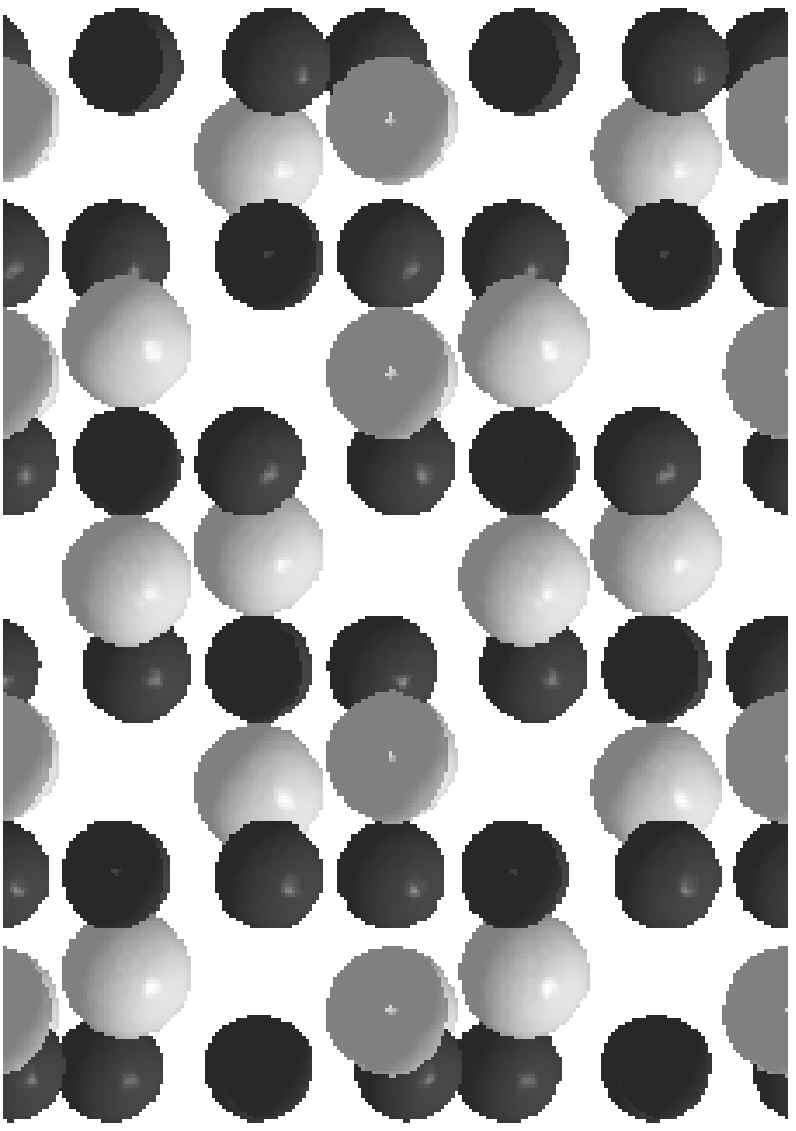}%
\raisebox{0.53in}{\includegraphics[scale=0.22]{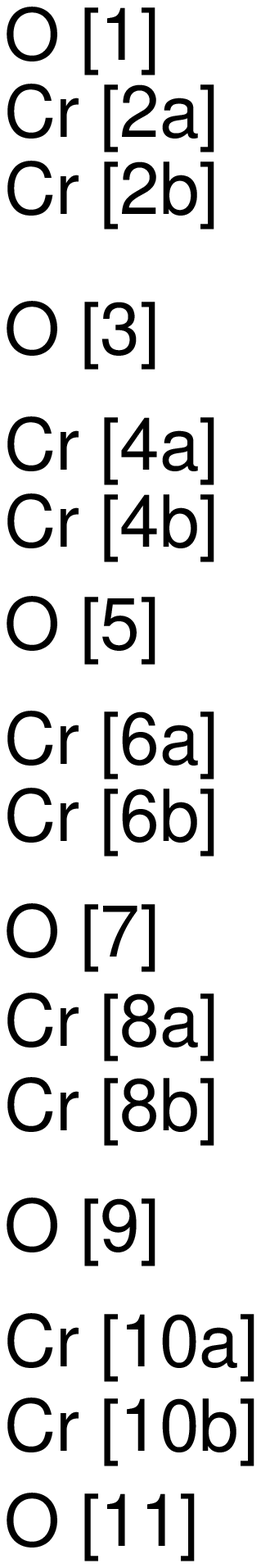}}%\rule[-0.4in]{0em}{0.4in}}%
\slabpic{b}{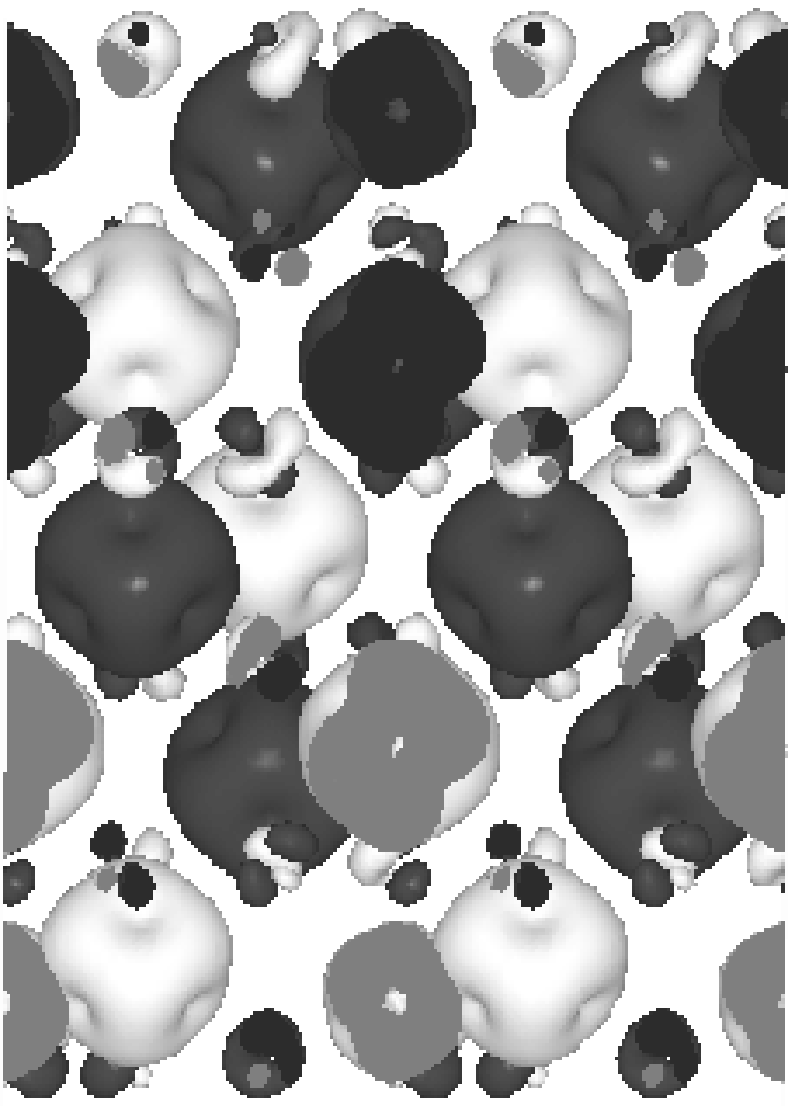}
\caption{ LSDA predictions for a twenty-eight atom surface slab of
\crtot{}: electron density (a) and spin polarization (b) (same
conventions as Figure~\ref{fig:spin_bulk_30}.) }
\label{fig:SurfaceChargeDensity} \label{fig:labelled_surf28}
\label{fig:spin_surface}
\end{center}
\end{figure}

Comparing the relaxed surface slab to the bulk, we find that the
primary structural change is a significant motion of surface oxygen
atoms (layer O[1] in Figure~\ref{fig:SurfaceChargeDensity}a) toward
the outermost sublayer of chromium atoms (layer Cr[2a]).  This motion
involves both a vertical component perpendicular to the surface and a
lateral component in the surface plane.  The vertical component
reduces the inter-planar spacing of layers O[1] and Cr[2a] by 33\%.
The lateral component rotates triangles of oxygen atoms centered above
the chromium atoms of layer Cr[2b] by 6.5\DEG\@, similar to the
rotation of 10\DEG\@ reported for \fetot{}~\cite{dft:fe2o3_scheffler}.
The second most significant change is an increase of 9\%, or
0.12~\AA{}, in the distance between the chromium layer Cr[2b] and the
next deeper oxygen layer O[3].  All shifts deeper within the structure
are less than 0.06~\AA\@.  Table~\ref{table:surf_pos_30} summarizes
our results.

The magnetic structure of the surface is significantly different from
that of the bulk (Figure~\ref{fig:spin_surface}b).  The most striking
feature is that the outermost chromium bilayer is now
ferromagnetically ordered.  In addition, we find a noticeable spin
moment on the outermost {\em oxygen} layer in the opposite spin
orientation, similar to what has been found in
\fetot{}\cite{dft:fe2o3_scheffler}.  From the figure, it is clear that
the origin of the net spin of the outer oxygen atoms is the
de-population of the lobes of opposing spin, which are normally filled
in the bulk.

\begin{figure}
\begin{center}
\slabpic{a}{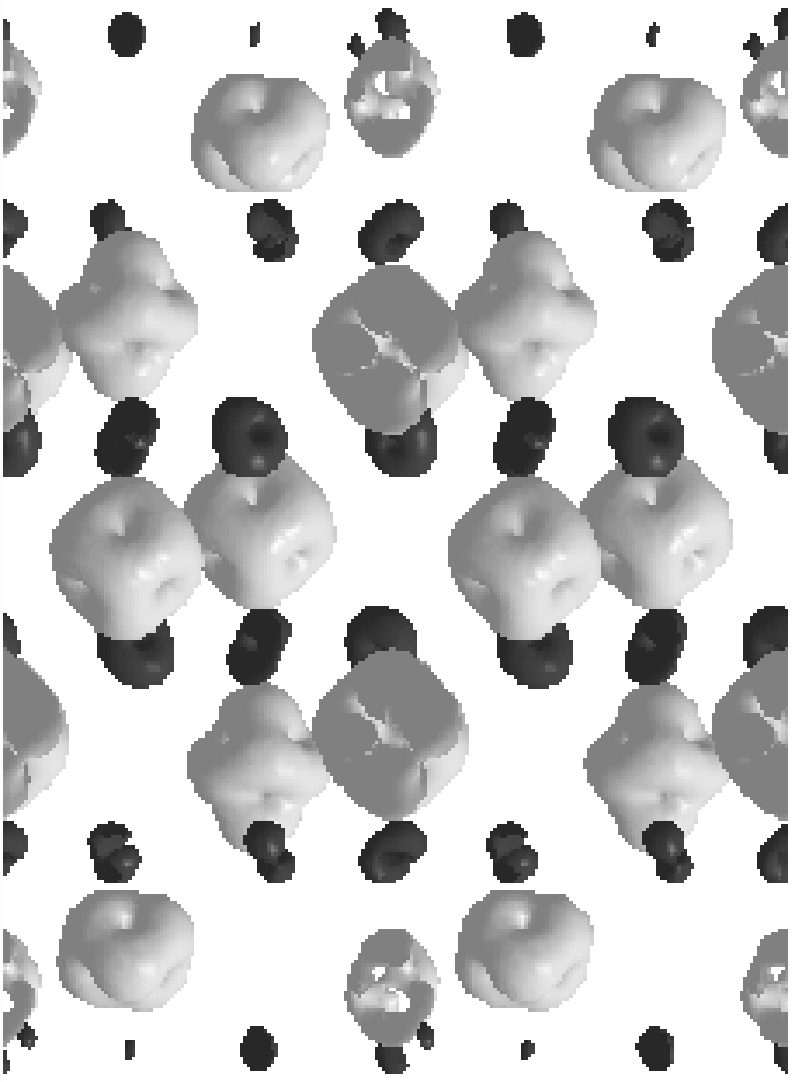}
\hspace{1ex}
\slabpic{b}{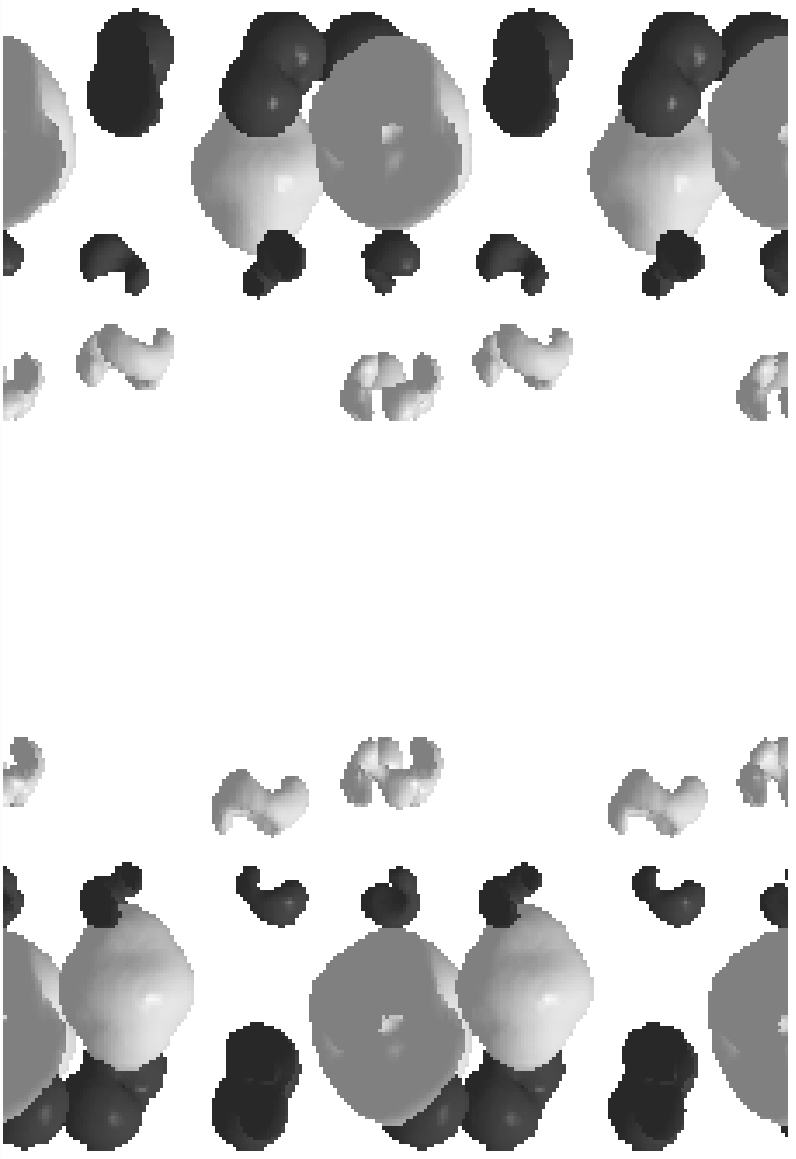}
\end{center}
\caption{Local chemical softness map for the surface slab of \crtot{}:
HOMOs (a), LUMOs (b).  Colors indicate proximity to chromium (white)
and oxygen (charcoal).}
\label{fig:homolumoox}
\label{fig:homolumored}
\end{figure}

In order to understand the chemical reactivity of the \crtot{}
surface, we compare the HOMO and LUMO states of the surface slab in
Figure~\ref{fig:homolumoox}a and \ref{fig:homolumoox}b.  These figures
show a remarkable spatial separation between the HOMOs and LUMOs.  The
HOMO states, which would interact most strongly with
electron-accepting species such as protons, contract into the bulk and
leave almost no concentration on the outer oxygen atoms.  On the other
hand, the LUMO states, which interact with electron-donating species
such as chloride or sulfide ions, concentrate strongly on the surface.
This suggests that such electron-donating species would be more
disruptive to the surface than would be their electron-accepting
counterparts.

\begin{figure}
\begin{center}
\slabpic{a}{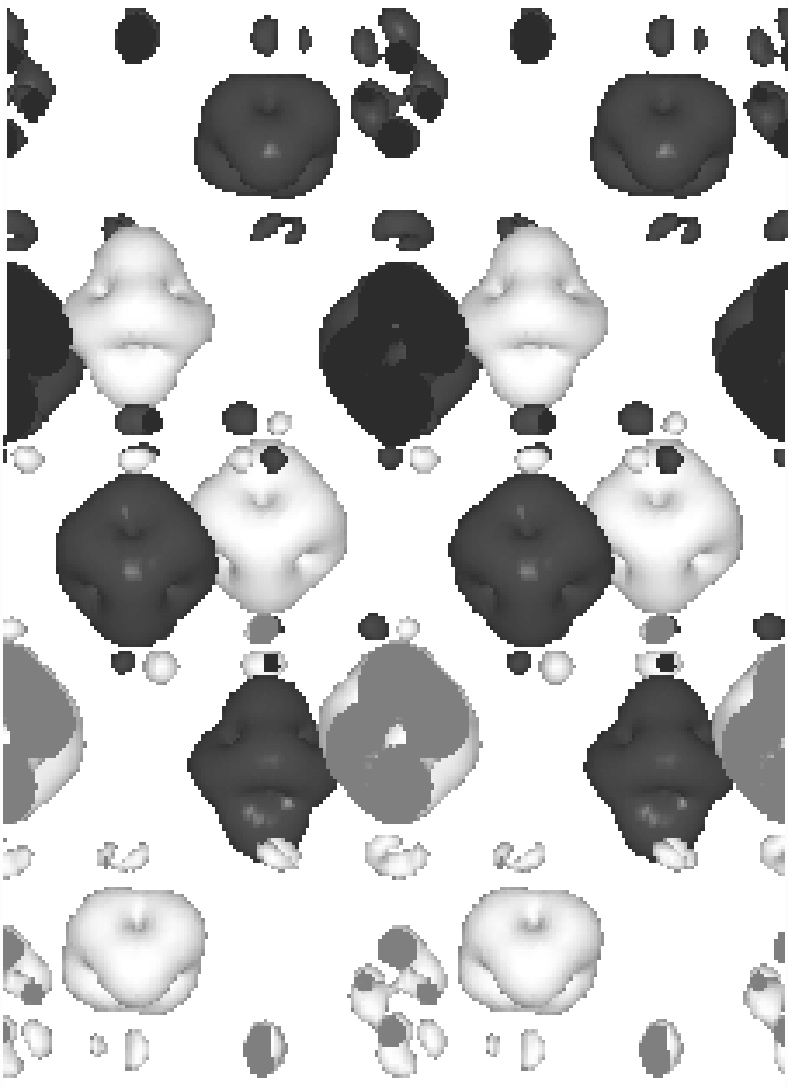}
\hspace{1ex}
\slabpic{b}{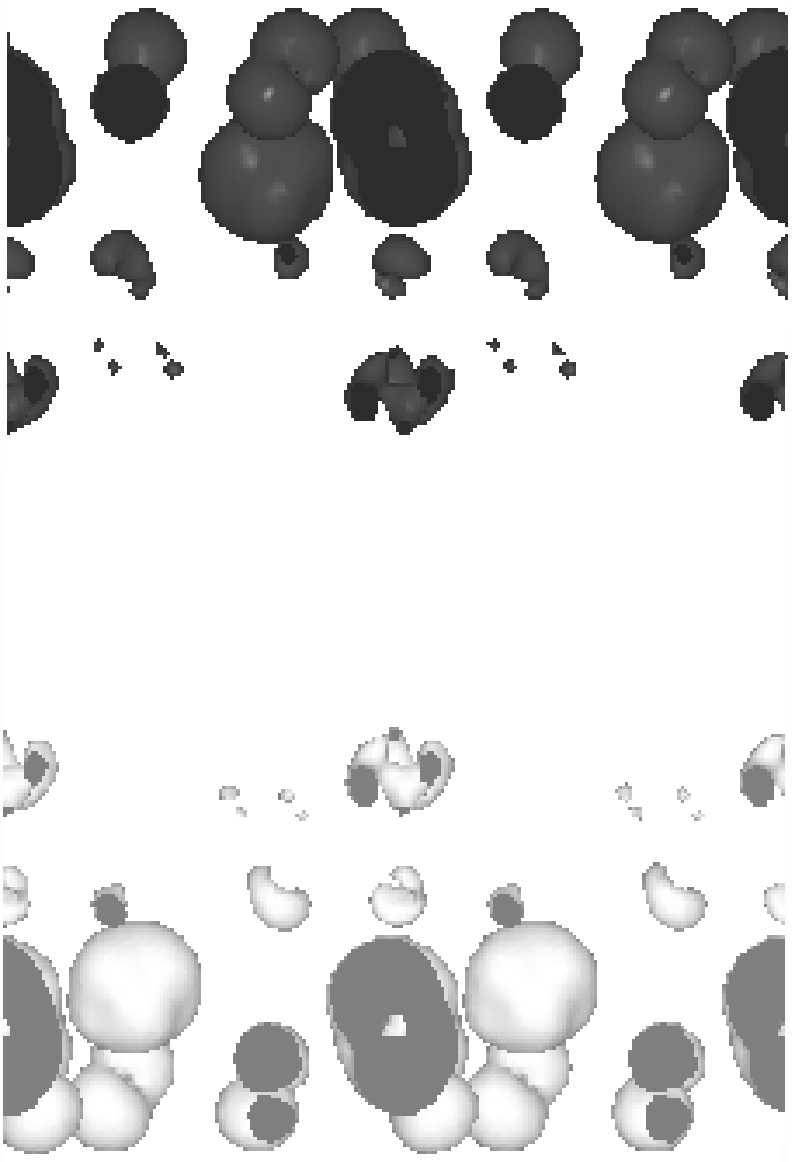}
\end{center}
\caption{Spin-dependence of the local chemical softness map for the
surface slab of \crtot{}: HOMOs (a), LUMOs (b). Colors indicate spin
direction.}
\label{fig:homolumo_spin_28}
\end{figure}

The spin polarization of the HOMOs and LUMOs on the chromium atoms
mimics the ferromagnetic behavior of the total spin density, as one
would expect from Hund's rule.  (See
Figure~\ref{fig:homolumo_spin_28}.)  However, the HOMO and LUMO states
of the oxygen atoms spin polarize oppositely to the total spin
density.  The reactive states of the surface are thus uniformly spin
polarized and provide an environment which encourages alignment of the
electronic spins of adsorbed species.  This provides an intriguing
starting point for understanding the conversion of para- to ortho-
hydrogen which has been observed to occur more readily in the presence
of magnetically ordered surfaces of \crtot{}\cite{AriasSelwood}.

\section{Conclusions}

The preceding results for the bulk crystal establish the critical role
of spin in the physics and chemistry of \crtot{}.  We show that the
plane-wave pseudopotential LSDA approach gives a good description of
the structure, mechanical response and magnetic order of the bulk.
Through this approach, we provide the first {\em ab initio} estimate
of the antiferromagnetic coupling constant.

The oxygen-terminated (0001) surface of \crtot{} exhibits strong
inward relaxation of the outermost oxygen layer with an associated
rotational reconstruction, as has been found in previous theoretical
studies of similar materials.  Understanding surface processes such as
catalysis and corrosion initiation, however, requires fundamental
knowledge of the electronic structure and cannot be inferred from the
surface relaxation alone.

We find that the electronic structure of the surface differs radically
from that of the bulk.  In particular, the outermost chromium bilayer
does not exhibit the antiferromagnetic order of the bulk but is
ordered ferromagnetically.  Moreover, the outer oxygen layer, rather
than exhibiting zero net spin polarization, develops an appreciable
net spin, but in the sense opposite to that of the underlying
ferromagnetic chromium bilayer.  Additionally, the surface manifests a
high density of low-energy unoccupied electronic states available to
influence chemical reactions at both oxygen and chromium surface
sites, but with negligible penetration into the bulk.  There is also some
density of high-energy occupied states reaching outward from the bulk
into the outermost chromium bilayer.  Remarkably, although the spins
of the outermost oxygen and chromium atoms orient in opposite
directions, the chemically relevant electronic states create a
blanketing environment of uniform spin which is available to encourage
spin alignment of adsorbed species.  This new understanding provides a
path for insight into novel magneto-chemical effects and their
implications for catalysis.

%%%%%%%%%%%%%%%%%%%%%%%%%%%%%%%%%%%%%%%%%%%%%%%%%%%%%%%%%%%%%%%%%%%%%%%%%%%%%%%

\begin{acknowledgements}
This work is partially supported by the National Computational Science
Alliance under Proposal \#DMR980011N and utilizes the Boston
University SGI/CRAY Origin2000\@.  J.A.C.\ gratefully acknowledges
funding provided by the Army Research Office under the University
Research Initiative (Projects 30345-CH-URI and 34173-CH-AAS, grant
numbers DAAL03-92-G-0177 and DAAH04-95-1-0302)\@.  A.A.R.\ was
partially funded by a Merrimack College Faculty Development Grant.
Thanks also to J.~W. Tester, R.~M. Latanision, D.~B. Mitton,
M.~T. Reagan, P.~A. Marrone, S.~Ismail-Beigi, D.~Yesilleten for their
help in a wide variety of capacities, and to C.~N. Hill and the MIT
Lab for Computer Science for use of the Pleiades Alpha-cluster.
\end{acknowledgements}

\bibliography{paper}
\bibliographystyle{unsrt}

\clearpage

\begin{table}
\begin{center}
\begin{tabular}{cccccc} %{|r|r@{ $\pm$ }l|}
Model               & $B$ (Mbar)      & $V$ (\AA$^3$)      & $E$ (eV)& 1/3 - $x$(O)  & $z$(Cr) - 1/3 \\ \hline
LDA                 & 3.39(10)      & 43.34(04)        & 23.63 & -0.004(2)   & 0.00626(1)  \\
LSDA                & 2.35(12)      & 47.19(14)        & 25.24 &  0.03133(8) & 0.01647(1)  \\
% LSDA(30)                & ---           & ---              & 25.49 &  0.02212(16) & 0.01196(4)  \\
UHF                 & 2.66(10)      & 50.5             & ---   & 0.03273     & 0.01722 \\ \hline
\multicolumn{6}{c}{\rule{0em}{1.2em}Published Experimental Values} \\ \hline
Ref.~\cite{bulk1}   & 2.38(12)      & 47.997(5)        & ---   & ---          & ---  \\
Ref.~\cite{bulk2}   & 2.31(30)      & 48.0             & ---   & 0.0282       & 0.01437 \\
Ref.~\cite{sherman} & ---           & ---              & 26.87$^\dagger$ & ---          & --- \\
Ref.~\cite{sawada}  & ---           & ---              & ---   & 0.02763      & 0.01417\\
\end{tabular}
${}^\dagger$ Computed with Born-Haber cycle. (See Text.)
\end{center}
\caption{ Comparison of computed bulk moduli $B$, volume per formula
unit $V$, cohesive energy $E$, and internal cell coordinates $x$ and
$z$ in ten-atom cell of \crtot{} with experimental values.  Volume and
cohesive energies (relative to isolated atoms) are per \crtot{}
formula unit.  }
\label{table:bulk_mod}
\end{table}

\begin{table}
\begin{center}
\begin{tabular}{c|c|c}
\textbf{Layer pair}  & \textbf{Bulk} &  \textbf{Surface} \\
 & \textbf{spacing [\AA{}]}  & \textbf{spacing [\AA{}]}  \\ \hline
O[1]--Cr[2a]  & 0.94 & 0.63  \\
Cr[2a]--Cr[2b]& 0.39 & 0.40  \\
Cr[2b]--O[3]  & 0.94 & 1.05  \\
O[3]--Cr[2b]  & 0.94 & 0.92  \\
\end{tabular}
\caption{
\label{table:surf_pos_30} \label{table:relaxdeltas}
{\em Ab initio} prediction of changes in interlayer spacing at the (0001) $\alpha$-\crtot{} surface.}
\end{center}
\end{table}

\end{document}